\documentclass{article}

\usepackage{graphicx}
\usepackage{float}
\usepackage[
singlelinecheck=false 
]{caption}

\begin{document}
\title{\bf Bending of light in novel 4$D$ Gauss-Bonnet-de Sitter black holes by Rindler-Ishak method}
\author{{Mohaddese Heydari-Fard$^{1}$ \thanks{Electronic address: m\_heydarifard@sbu.ac.ir}, Malihe Heydari-Fard$^{2}$\thanks{Electronic address: heydarifard@qom.ac.ir} and Hamid Reza Sepangi$^{1}$\thanks{Electronic address: hr-sepangi@sbu.ac.ir}}\\ {\small \emph{$^{1}$ Department of Physics, Shahid Beheshti University, G. C., Evin, Tehran 19839, Iran}}
\\{\small \emph{$^{2}$ Department of Physics, The University of Qom, 3716146611, Qom, Iran}}}

\maketitle

\begin{abstract}
We study the bending of light in the space-time of black holes in four-dimensional Einstein-Gauss-Bonnet theory of gravity, recently proposed by Glavan and Lin \cite{Glavan}. Using Rindler-Ishak method, the effect of Gauss-Bonnet coupling on the bending angle is studied. We show that a positive Gauss-Bonnet coupling gives a negative contribution to the Schwarzschild-de Sitter deflection angle, as one would expect.
\vspace{5mm}
\vspace{1mm}
\end{abstract}

\section{Introduction}
Many efforts have gone into understanding the nature of dark energy as being responsible of accelerated expansion of the observed universe. As is well known, one of the prime candidates for dark energy  is the cosmological constant $\Lambda$ \cite{1}--\cite{6}. Study of the effects of cosmological constant on local phenomena such as null geodesics, time delay of light, gravitational time advancement and the perihelion precession, has attracted considerable attention in the last decades \cite{7}--\cite{ta5}. Many authors have investigated effects of the cosmological constant on bending of light. Islam was the first to mention that the null geodesic equation in a spherically symmetric space-time does not involve $\Lambda$ term and concluded that the cosmological constant does not affect the bending of light \cite{Islam}. However, later Rindler and Ishak \cite{Rindler}, by considering the intrinsic properties of the Schwarzschild-de Sitter space-time, proposed a new method for calculating the deflection angle of light. This approach gives a new insight into the effect of the cosmological constant on bending of light and, as they put it,  ``$\Lambda$ does contribute to the observed bending of light.'' Since the cosmological constant effectively counteracts gravity, one may intuitively expect that a positive $\Lambda$ would diminish the bending of light. The different perspectives of the method such as the integration of gravitational potentials and Fermat’s principle have been studied in \cite{Ishak2}. The studies of light bending in Kerr-de Sitter and  Reissner-Nordstrom-de Sitter space-times, using  Rindler-Ishak method, have been carried out in \cite{Kerr} and \cite{RN}, respectively. Also the method has been applied to galactic halos, of which an excellent example is the Mannheim-Kazanas solution of conformal Weyl gravity \cite{Weyl1}--\cite{Weyl3}. For further insight into the use of Rindler-Ishak method, see \cite{T1}--\cite{T5}.

The low energy limit of string theory contains quadratic and higher-order curvature invariants in addition to the usual scalar curvature in the Lagrangian. One specific combination of higher-order curvature invariants is the Gauss-Bonnet term
\begin{equation}
{\cal {G}}=R^2-4R_{\mu\nu}R^{\mu\nu}+R_{\mu\nu\rho\sigma}R^{\mu\nu\rho\sigma},
\label{b}
\end{equation}
which is a natural extension of Einstein’s general relativity in a $D$-dimensional space-time with $D-4$ extra dimensions. In $D=4$ the Gauss-Bonnet term is a topological invariant and does not contribute to the gravitational field equations. However, recently a four-dimensional Einstein-Gauss-Bonnet (EGB) gravity has been proposed by Glavan and Lin \cite{Glavan} where, by re-scaling the coupling constant $\alpha\rightarrow\frac{\alpha}{D-4}$, and in the limit $D\rightarrow 4$, the GB term does contribute to the gravitational dynamics, thus circumventing the conditions of Lovelock’s theorem. The theory preserves the number of degrees of freedom and avoids the Ostrogradsky instability. Also, it has been argued \cite{Glavan} that the gravitational force is repulsive at short distances and thus no particle can reach the curvature singularity at $r=0$. However, by considering the radial ingoing geodesics of a free-falling massive particle, the authors in \cite{Arrechea} have explicitly proved that such a claim is not correct and the curvature singularity can be reached within a finite proper time. It should be noted that such a 4$D$ black hole solution was initially found in semi-classical gravity with a conformal anomaly \cite{Cai}. However, the present theory can be considered as a classical modified gravity on equal footing with general relativity.

There have been some criticisms concerning the regularization method used in \cite{Glavan}. It has been argued that taking the $D\rightarrow 4$ limit may not be consistent and the theory and thus is not well defined in four dimensions \cite{c1}--\cite{c7}. To address this problem, some prescriptions such as compactifying the $D$-dimensional EGB gravity \cite{n1}--\cite{n2}, adding a counter term to the action \cite{n3}--\cite{n4} and breaking the temporal diffeomorphism invariance \cite{n5} have been proposed to obtain a consistent theory of 4$D$ EGB gravity. It is worth stressing that in these consistent theories the spherically symmetric black hole solution of \cite{Glavan} is also valid and has been extensively studied from many angles. For instance, black hole solutions and their extensions to charged and rotating black holes were studied and carried out in \cite{charge} and \cite{rotating}, respectively. In addition, a general study of  solutions in Lovelock gravity has been presented in \cite{love1}--\cite{love2}. Also, an exact charged black hole solution of the theory surrounded by clouds of strings was obtained in \cite{string}. The non-static solutions of radiating black holes as well as the Hayward and Bardeen black hole solutions were constructed in \cite{rad1}--\cite{bardeen}, respectively. Also, there has been many studies in the analysis of some properties in Bardeen black holes including strong gravitational lensing \cite{ba1}, bound orbits and circular geodesics \cite{ba2}--\cite{ba4}. Exact spherically symmetric wormhole solutions for an isotropic and anisotropic matter source and thin-shell wormholes  were constructed in \cite{w1} and \cite{w2}. The authors in \cite{star} have investigated the structure of relativistic stars and discussed observational constraints on the GB coupling $\alpha$. The study of geodesic motion and the effects of GB coupling on shadows cast by black holes in four-dimensional EGB gravity have been considered in \cite{isco} and \cite{s}. Also, quasinormal modes of black holes in the framework of EGB gravity have been studied in \cite{QNM}. In the strong deflection limit,  gravitational lensing by static and spherically symmetric black holes has been investigated in \cite{lensing}. The analysis of gravitational instability of asymptotically flat, de-Sitter and anti-de Sitter black holes in four-dimensional EGB gravity was carried out in \cite{stability}. Also, phase transition and thermodynamical behavior of black holes have been investigated in \cite{th1}--\cite{th3}. For other studies on four-dimensional EGB gravity see \cite{a1}--\cite{a9}.

The purpose of this note is to investigate the problem of light bending in  EGB black holes by the Rindler-Ishak method. We investigate possible corrections resulting from GB coupling $\alpha$ on the deflection angle of light.

\section{Bending angle in 4$D$ Einstein-Gauss-Bonnet-de Sitter black hole}
Bending of light in the vicinity of compact gravitational objects is one of the most studied problems in general relativity and was first confirmed by Eddington and Dyson in 1919 \cite{Eddington}. The action of 4$D$ EGB gravity is given by
\begin{equation}
S=\frac{1}{16\pi G}\int\sqrt{-g}d^4x\left[R+\alpha\left(R^2-4R_{\mu\nu}R^{\mu\nu}+R_{\mu\nu\lambda\rho}R^{\mu\nu\lambda\rho}\right)\right],
\label{1}
\end{equation}
where $\alpha$ is the GB coupling with dimensions of length squared. The static and spherically symmetric solution of the theory is given by \cite{Glavan}
\begin{equation}
ds^2=-f(r)dt^2+\frac{dr^2}{f(r)}+r^2(d\theta^2+\sin^2\theta d\phi^2),
\label{2}
\end{equation}
where
\begin{equation}
f(r)=1+\frac{r^2}{2\alpha}\left(1\pm\sqrt{1+\frac{8\alpha M}{r^3}}\right),
\label{3}
\end{equation}
and  $M$ denotes mass of the black hole. As is mentioned in \cite{Glavan}, when $\alpha$ takes negative values there is no real solution at short radial distances for which $r^3<-8\alpha M$. However, for positive values of $\alpha$ there are two branches of solutions. The negative branch asymptotically behaves as Schwarzschild solution with positive mass sign, whereas the positive branch behaves as Schwarzschild-de Sitter solution with a negative gravitational mass. Since the universe is in an accelerating expansion phase and the cosmological constant is the simplest candidate for its explanation, it would be interesting to study the bending of light in the presence of a $\Lambda$-term in the theory. In this case the metric function $f(r)$ reads  \cite{charge}
\begin{equation}
f(r)=1+\frac{r^2}{2\alpha}\left(1\pm\sqrt{1+\frac{8\alpha M}{r^3}+\frac{4\alpha\Lambda}{3}}\right),
\label{100}
\end{equation}
where the negative branch asymptotically goes to the Schwarzschild-de Sitter solution.

Now, let us obtain the deflection angle of light in EGB gravity. The standard approach for calculating the bending angle is \cite{winberg}
\begin{equation}
\Delta\phi =2|\phi(\infty)-\phi(r_0)|-\pi,
\label{a101}
\end{equation}
where $r_0$ is the closest distance to the black hole. However, the space-time we consider here is not asymptotically flat and $r\rightarrow \infty$ does not make sense. Therefore, we use Rindler-Ishak method proposed in \cite{Rindler} to obtain  the deflection angle in an asymptotically non-flat space-time. Although the null geodesic equation for Schwarzschild-de Sitter space-time does not involve the $\Lambda$ term, Rindler and Ishak have shown that by considering the effects of cosmological constant on the geometry of space-time one can obtain the contribution of $\Lambda$ to the bending angle. Using the Euler-Lagrange equations for null geodesics in  equatorial plane, $\theta =\frac{\pi}{2}$, we obtain the following equation
\begin{equation}
\frac{d^2u}{d\phi^2}+u\approx3Mu^2-2M\alpha\Lambda u^2-12\alpha M^2u^5,
\label{5}
\end{equation}
where we have only kept the terms to first order in $\alpha$ and $\Lambda$  and the variable $u$ is defined as $u(\phi)=\frac{1}{r}$. As one can see, the second term in the above equation involves both the cosmological constant and GB coupling, namely in this theory the orbital equation of light contains the cosmological constant in contrast to Schwarzschild-de Sitter space-time in general relativity. Moreover, we note that terms $3Mu^2$, $12\alpha M^2 u^5$ and $2M\alpha\Lambda u^2$ are much smaller than the term $u$. For instance, it is easy to see that for a compact object like the Sun, the ratio $\frac{3Mu^2}{u}=\frac{3R_{\rm S}}{R_{\odot}}$ is of the order of $10^{-6}$, with $R_{\rm S}$ being the Schwarzschild radius. On this assumptions, we solve equation (\ref{5}) using a perturbative method up to the third order and consider a solution as
\begin{equation}
u=u_0+\delta u_1+\delta u_2+\delta u_3+\cdots,
\label{bb}
\end{equation}
where $u_0=\frac{\sin\phi}{R}$ is the un-deflected straight line path and $R$ is the shortest distance between the mass and un-deflected light rays in flat space, see figure 1. The corrections $\delta u_1$, $\delta u_2$ and $\delta u_3$ respectively satisfy the following equations
\begin{eqnarray}
\frac{d^2(\delta u_1)}{d\phi^2}+\delta u_1=3 Mu_0^2,
\label{n1}
\end{eqnarray}
\begin{eqnarray}
\frac{d^2(\delta u_2)}{d\phi^2}+\delta u_2=6 Mu_0\delta u_1,
\label{n2}
\end{eqnarray}
\begin{eqnarray}
\frac{d^2(\delta u_3)}{d\phi^2}+\delta u_3=6 Mu_0\delta u_2+3 M\delta u_1^2-2M\alpha\Lambda u_0^2-12\alpha M^2u_0^5.
\label{n3}
\end{eqnarray}
\begin{figure}
\centering
\includegraphics[width=3.0in]{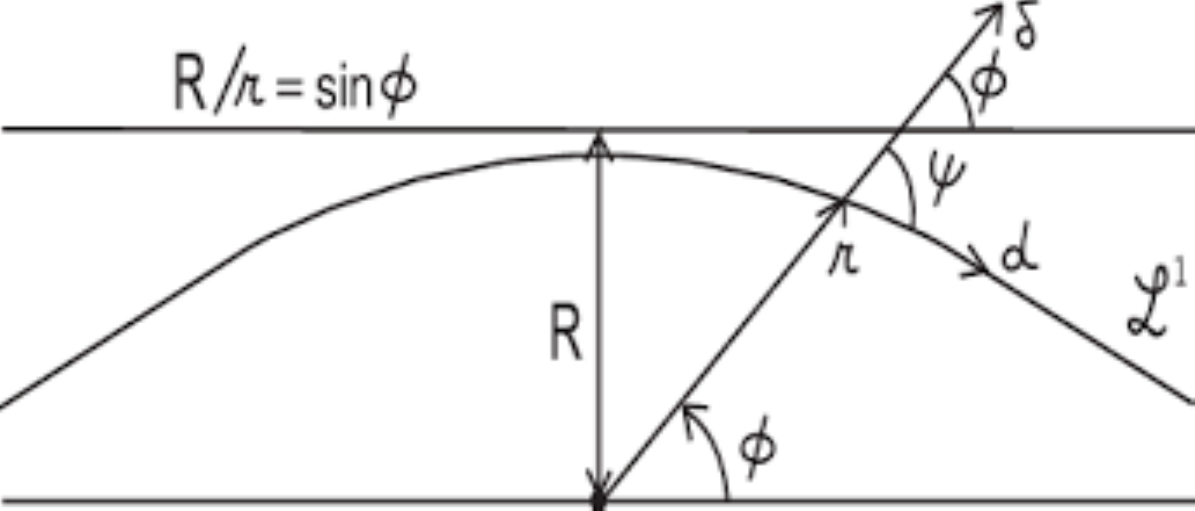}
\caption{Light bending in the space-time of a black hole. The one-sided
deflection angle is $\psi-\phi\equiv\epsilon$. The figure is taken from \cite{Rindler}.}
\label{Planetary}
\end{figure}
Plugging the solutions of (\ref{n1})-(\ref{n3}) into equation (\ref{bb}) we obtain
\begin{eqnarray}
u(\phi)&=&\frac{\sin\phi}{R}+\frac{3M}{2R^2}\left(1+\frac{1}{3}\cos2\phi\right)+\frac{3M^2}{16R^3}\left(-\sin3\phi+10(\pi-2\phi)\cos\phi\right)\nonumber\\
&-&\frac{M\alpha\Lambda}{R^2}\left(1+\frac{1}{3}\cos2\phi\right)+\frac{M^2\alpha}{32R^5}\left(\sin5\phi-15\sin3\phi-60(\pi-2\phi)\cos\phi\right)\nonumber\\
&+&\frac{M^3}{16R^4}\left(195-\cos4\phi-30(\pi-2\phi)\sin2\phi+86\cos2\phi\right).
\label{66}
\end{eqnarray}
Now, following Rindler-Ishak, we compute angle $\psi$ between the photon orbit direction $\bf{d}$ and  direction of $\phi=$const line, by the invariant formula
\begin{equation}
\cos\psi=\frac{g_{ij}d^{i}\delta^{j}}{\sqrt{g_{ij}d^{i}d^{j}}\sqrt{g_{ij}\delta^{i}\delta^{j}}},
\label{7}
\end{equation}
where $g_{ij}$ are the coefficients of the 2-metric on $\theta=\frac{\pi}{2}$, $t=$const. surface. Substituting $d=(dr,d\phi)$ and $\delta=(\delta r,0)$ in equation (\ref{7}) we find
\begin{equation}
\cos\psi=\frac{|dr/d\phi|}{\sqrt{|dr/d\phi|^2+f(r)r^2}},
\label{8}
\end{equation}
or equivalently
\begin{equation}
\tan\psi=\frac{r\sqrt{f(r)}}{|dr/d\phi|}.
\label{9}
\end{equation}
To obtain the one-sided deflection angle at the point where $\phi\ll 1$ we have
\begin{eqnarray}
u=\frac{1}{r}\approx\frac{2M}{R^2}\left(1+\frac{\phi R}{2M}+\frac{3M}{32R}(10\pi-23\phi)+\frac{M^2}{32R^2}(280-60\pi\phi)-\frac{2\alpha\Lambda}{3}+\frac{M\alpha}{46R^3}(80\phi-60\pi)\right),
\label{10}
\end{eqnarray}
and
\begin{eqnarray}
\frac{dr}{d\phi}=-\frac{r^2}{R}\left(1-\frac{69M^2}{16R^2}-\frac{15\pi M^3}{4R^3}+\frac{5M^2\alpha}{2R^4}\right).
\label{11}
\end{eqnarray}
Finally by substituting in equation (\ref{9}), we find the following expression for the total deflection angle
\begin{eqnarray}
2\epsilon=2(\psi-\phi)&\approx&\left[\frac{4M}{R}+\frac{15\pi M^2}{4R^2}+\frac{177M^3}{4R^3}-\frac{\Lambda R^3}{6M}+\frac{25\pi\Lambda M^2}{512}
-\frac{71R\Lambda M}{96}+\frac{5\pi\Lambda R^2}{32}\right]\nonumber\\
&-&\alpha\left[\frac{15\pi M^2}{4R^4}+\frac{9\Lambda M}{4R}+\frac{5\pi\Lambda}{32}+\frac{R^3\Lambda^2}{18M}+\frac{5\pi R^2\Lambda^2}{96}\right]
.\label{d1}
\end{eqnarray}

As one can see, the deflection angle is modified by new terms containing the GB coupling and cosmological constant. The first term is the standard one for bending of light in a Schwarzschild field. In addition to the usual term $-\frac{\Lambda R^3}{6M}$ for the Schwarzschild-de Sitter space-time obtained initially in \cite{Rindler}, the above expression contains the term $-\frac{71R\Lambda M}{96}$ similar to that obtained by Sereno which couples the cosmological constant to the mass of the lens and coming from the higher order corrections \cite{Sereno}. Also we obtain a local term $\frac{25\pi\Lambda M^2}{512}$ which does not depend on $R$. Since all terms in the second bracket are positive, we find that for a positive $\alpha$, the effect of the GB coupling is to diminish the deflection angle.

In order to get an expression for the bending angle in terms of the distance of the closest approach $r_0$ which is the value of $r$ when $\phi=\frac{\pi}{2}$, we find the following relation from equation (\ref{66}) in the absence of $\alpha$ and $\Lambda$ parameters
\begin{equation}\label{300}
\frac{1}{r_0}=\frac{1}{R}+\frac{M}{R^2}+\frac{3M^2}{16R^3}+\frac{27M^3}{4R^4}
\end{equation}
which relate $R$ to the distance of closest approach $r_0$. Thus the deflection angle, up to second order in $\frac{M}{r_0}$, is given by
\begin{equation}
2\epsilon\approx\frac{4M}{r_0}+\left(\frac{15\pi}{4}-4\right)\frac{M^2}{r_0^2},\label{301}
\end{equation}
which is in agreement with equation (19) of \cite{Will} for the asymptotically flat Schwarzschild space-time.

The effect of the cosmological constant on deflection angle at small scales such as the solar system is expected to be negligible. Therefore by setting $\Lambda =0$ in equation (\ref{d1}) and keeping the linear terms in the expansion, we find
\begin{equation}
2\epsilon\simeq\frac{4M}{R}\left(1-\frac{15\pi\alpha M}{16R^3}\right)
,\label{f}
\end{equation}
which only gives corrections of the GB term to the deflection angle. The observational data on light deflection by the Sun, from long baseline radio interferometry \cite{sun}, gives $\delta\phi_{LD}=\delta\phi_{LD}^{(GR)}(1+\Delta_{LD})$, with $\Delta_{LD}\leq0.0002\pm0.0008$, where $\delta\phi_{LD}^{(GR)}=1.7510$ arcsec. By assuming that $\Delta_{LD}$ is entirely due to the geometrical effects of the GB term, the observational results constrain the coupling constant $\alpha$ to $|\alpha|\leq\frac{16R^3\Delta_{LD}}{15\pi M}$. Taking for $R$ and $M$ the values of radius and mass of the Sun, $R_{\odot}=6.955\times10^{8}\rm m$ and $M_{\odot}=1.989\times10^{30}\rm kg$, we have $|\alpha|\leq(1.55\pm6.19)\times10^{19}\rm m^{2}$. This result is compatible with what was obtained in \cite{solar} by solar system measurements on GB gravity. Nevertheless, a better way for comparing theoretical results with experimental data is to use the post-Newtonian (PN) approach. This is a successful formalism for the study of classical tests in modified gravity theories where the metric of the theory is obtained in the weak field limit and its departure from general relativity is expressed in terms of the PN parameters \cite{PN1}--\cite{PN2}. Then, by calculating the theoretical predictions of classical tests such as  the bending of light which depend on the PN parameters and comparing them with observational data, one obtains the constraints on the parameters of the modified theories of gravity.

Finally, it is worth mentioning that although the purpose of the present paper is to investigate the problem of light bending in an asymptotically non-flat space-time using the Rindler-Ishak method, one can use other classical tests in the solar system such as the gravitational time delay, the Cassini tracking experiment and the perihelion shift of the planets, to constrain both the cosmological constant and the GB coupling. So, in what follows we briefly study the problem of planet’s perihelion advance in the space-time described by the metric function (\ref{100}) to obtain an upper bound on the cosmological constant $\Lambda$ and coupling parameter $\alpha$.

The equation of motion for massive particles in this space-time can be obtain as
\begin{equation}
\frac{d^2u}{d\phi^2}+u=\frac{M}{L^2}+\left(3M-2M\alpha\Lambda\right)u^2-\frac{8\alpha M^2}{L^2}u^3-12\alpha M^2u^5+\left(\frac{\alpha\Lambda ^2}{9L^2}-\frac{\Lambda}{3L^2}\right)\frac{1}{u^3},
\label{p1}
\end{equation}
where $L$ is the conserved quantity along the particle trajectory. To first order, this equation has a solution of the form
\begin{equation}
u(\phi)\simeq\frac{M}{L^2}\left[1+e \cos \left(1-\frac{3M^2}{L^2}-\frac{\Lambda L^6}{2M^4}-\frac{12\alpha M^4}{L^6}-\frac{30\alpha M^6}{L^8}+\frac{2M^2\alpha\Lambda}{L^2}+\frac{\alpha\Lambda^2L^6}{6M^4}\right)\phi\right],
\label{p2}
\end{equation}
where $e$ is the eccentricity. Therefore, the perihelion shift after one revolution is given by
\begin{equation}
\Delta\phi\simeq2\pi\left[\frac{3M}{a(1-e^2)}+\frac{\Lambda a^3(1-e^2)^3}{2M}+\frac{12\alpha M}{a^3(1-e^2)^3}+\frac{30\alpha
M^2}{a^4(1-e^2)^4}-\frac{2M\alpha\Lambda}{a(1-e^2)}-\frac{\alpha\Lambda^2a^3(1-e^2)^3}{6M}\right].
\label{p3}
\end{equation}
Here, we have used $\frac{L^2}{M}=a(1-e^2)$, with $a$ being the semi-major axis. The first term, $\frac{6\pi M}{a(1-e^2)}$, is the standard value for the perihelion precession in the Schwarzschild space-time. The second term represents the contribution of the cosmological constant obtained earlier in \cite{landa}. The other terms contain modifications brought about as a result of the GB coupling. Using the difference between the general relativistic values of the perihelion shift and the observed values \cite{data}, we have $|\alpha|\leq2.02\times10^{16}\rm m^{2}$ which is three order of magnitude smaller than that given by the bending of light. Also, we obtain $|\Lambda|\leq2.81\times10^{-41}\rm m^{-2}$ which is compatible with what was obtained in \cite{ref1}.

\section{Conclusions}
In this paper we have studied the bending of light in the space-time of a four-dimensional EGB-de Sitter black hole using the method of Rindler and Ishak. We have obtained an expression for the bending angle which contains the effects of the GB coupling and cosmological constant. We found that for a positive coupling constant $\alpha$ the effect of GB term is to diminish the bending angle in the novel four-dimensional EGB gravity even up to third order in $M$. This is an interesting result in that the GB term, as a candidate for dark energy, counteracts gravity. Also using the observational data on the bending of light by the Sun and on the perihelion shift of the inner planets of the solar system, we obtained a constraint on the GB coupling.

\section*{Acknowledgements}
We would like to thank the anonymous referees for valuable and interesting comments.


\begin{thebibliography}{99}
\bibitem{1} S. Weinberg, {\it Rev. Mod. Phys} {\bf 61} (1989) 1.
\bibitem{2} M. S. Turner, {\it Phys. Rep} {\bf 619} (2000) 333.
\bibitem{3} V. Sahni and A. Starobinsky, {\it Int. J. Mod. Phys} D {\bf 09} (2000) 373.
\bibitem{4} T. Padmanabhan, {\it Phys. Rep}, {\bf 380} (2003) 235.
\bibitem{5} P. J. E. Peebles and B. Ratra, {\it Rev. Mod. Phys} {\bf 75} (2003) 559.
\bibitem{6} A. Upadhye, M. Ishak and P. J. Steinhardt, {\it Phys. Rev.} D {\bf 72} (2005) 063501.


\bibitem{7} B. Chen, R. Kantowski and X. Dai, {\it Phys. Rev.} D {\bf 82} (2010)  043005.
\bibitem{8} K. E. Boudjemaa, M. Guenouche and S. R. Zouzou, {\it Gen. Rel. Grav} {\bf 43} (2011) 1707.



\bibitem{9} R. A. Alpher, {\it Am. J. Phys} {\bf 35} (1967) 771.
\bibitem{10} A.W. Kerr, J. C. Hauck and B. Mashhoon, {\it Class. Quant. Grav} {\bf 20} (2003) 2727.
\bibitem{11} N. Cruz, M. Olivares and J. R. Villanueva, {\it Class. Quant. Grav} {\bf 22} (2005) 1167.
\bibitem{12} E. Hackmann and C. Lammerzahl, {\it Phys. Rev. Lett} {\bf 100} (2008) 171101.


\bibitem{ref1} Y. Xie and X. M. Deng, {\it Mon. Not. Roy. Astron. Soc} {\bf 433} (2013) 3584.
\bibitem{ref2} X. M. Deng and Y. Xie, {\it Eur. Phys. J.} C {\bf 75} (2015) 539.
\bibitem{ref3} S. S. Zhao and Y. Xie, {\it JCAP} {\bf 07} (2016) 007.
\bibitem{ref4} X. Lu, F. W. Yang and Y. Xie, {\it Eur. Phys. J.} C {\bf 76} (2016) 357.
\bibitem{ref5} X. M. Deng, {\it EPL} {\bf 120} (2017) 60004.
\bibitem{ref6} X. M. Deng and Y. Xie, {\it Phys. Lett.} B {\bf 772} (2017) 152.
\bibitem{ref7} X. M. Deng, {\it Eur. Phys. J.} Plus {\bf 132} (2017) 85.
\bibitem{ref8} W. G. Cao and Y. Xie, {\it Eur. Phys. J.} C {\bf 78} (2018) 191.
\bibitem{ref9} F. Y. Liu, Y. F. Mai, W. Y. Wu and Y. Xie, {\it Phys. Lett.} B {\bf 795} (2019) 475.
\bibitem{ref10} X. Lu and Y. Xie, {\it Eur. Phys. J.} C {\bf 80} (2020) 625.


\bibitem{ta1} A. Bhadra and K. K. Nandi, {\it Gen. Rel. Grav} {\bf 42} (2010) 293.
\bibitem{ta2} S. Ghosh and A. Bhadra, {\it Eur. Phys. J.} C {\bf 75} (2015) 494.
\bibitem{ta3} X. M. Deng, {\it Class. Quant. Grav} {\bf 35} (2018) 175013.
\bibitem{ta4} G. Li and X. M. Deng, {\it Commun. Theor. Phys} {\bf 70} (2018) 721.
\bibitem{ta5} X. M. Deng, {\it Mod. Phys. Lett} A {\bf 33} (2018) 1850110.


\bibitem{Islam} N. J. Islam, {\it Phys. Lett.} A {\bf 97} (1983) 239.


\bibitem{Rindler} W. Rindler and M. Ishak, {\it Phys. Rev.} D {\bf 76} (2007) 043006.


\bibitem{Kerr} J. Sultana, {\it Phys. Rev.} D {\bf 88} (2013) 042003.
\bibitem{RN} M. Heydari-Fard, S. Mojahed and S. Y. Rokni {\it Astrophys Space Sci} {\bf 351} (2014) 251.


\bibitem{Ishak2} M. Ishak, {\it Phys. Rev.} D {\bf 78} (2008) 103006.


\bibitem{Weyl1} A. Bhattacharya, R. Isaev, M. Scalia, C. Cattani and K. K. Nandi, {\it JCAP} {\bf 09} (2010) 004.
\bibitem{Weyl2} A. Bhattacharya, G. M. Garipova, E. Laserra, A. Bhadra and K. K. Nandia, {\it JCAP} {\bf 02} (2011) 028.
\bibitem{Weyl3} C. Cattani, M. Scalia, E. Laserra, I. Bochicchio and K. K. Nandi, {\it Phys. Rev.} D {\bf 87} (2013) 047503.


\bibitem{T1} G. Farrugia, J. L. Said and M. L. Ruggiero, {\it Phys. Rev.} D {\bf 93} (2016) 104034.
\bibitem{T2} A. Mishra and S. Chakraborty, {\it Eur. Phys. J.} C {\bf 78} (2018) 374.
\bibitem{T3} M. Guenouche and S. R. Zouou, {\it Phys. Rev.} D {\bf 98} (2018) 123508.
\bibitem{T4} M. S. Ali and S. Bhattacharya, {\it Phys. Rev.} D {\bf 97} (2018) 024029.
\bibitem{T5} M. Haluk Secuk and O. Delice, {\it Eur. Phys. J.} C {\bf 135} (2020) 610.


\bibitem{Glavan} D. Glavan and C. Lin, {\it Phys. Rev. Lett} {\bf 124} (2020) 081301.


\bibitem{Arrechea} J. Arrechea, A. Delhom and J. Cano, {\it Chin. Phys.} C {\bf 45} (2021) 013107.


\bibitem{Cai} R. G. Cai, L. M. Cao and N. Ohta, {\it JHEP} {\bf 1004} (2010) 082; R. G. Cai, {\it Phys. Lett.} B, {\bf 733} (2014) 183.


\bibitem{c1} W. Y. Ai, {\it Commun. Theor. Phys} {\bf 72} (2020) 095402.
\bibitem{c2} M. Gurses, T. C. Sisman and B. Tekin, {\it Eur. Phys. J.} C {\bf 80} (2020) 647.
\bibitem{c3} S. Mahapatra, {\it Eur. Phys. J.} C {\bf 80} (2020) 992.
\bibitem{c4} F. W. Shu, {\it Phys. Lett.} B {\bf 811} (2020) 135907.
\bibitem{c5} S. X. Tian and Z. H. Zhu, arXiv:2004.09954 [gr-qc].
\bibitem{c6} J. Bonifacio, K. Hinterbichler and L. A. Johnson, {\it Phys. Rev.} D {\bf 102} (2020) 024029.
\bibitem{c7} J. Arrechea, A. Delhom and A. Jim\'{e}nez-Cano, {\it Chin. Phys.} C {\bf 45} (2021) 013107.


\bibitem{n1} H. L\"{u} and Y. Pang, {\it Phys. Lett.} B {\bf 809} (2020) 135717.
\bibitem{n2} T. Kobayashi, {\it JCAP} {\bf 07} (2020) 013.


\bibitem{n3} P. G. S. Fernandes, P. Carrilho, T. Clifton and D. J. Mulryne, {\it Phys. Rev.} D {\bf 102} (2020) 024025.
\bibitem{n4} R. A. Hennigar, D. Kubiz\v{n}\'{a}k, R. B. Mann and C. Pollack, {\it JHEP} {\bf 2020} (2020) 27.


\bibitem{n5} K. Aoki, M. A. Gorji and S. Mukohyama, {\it Phys. Lett.} B {\bf 810} (2020) 135843.


\bibitem{charge} P. G. S. Fernandes, {\it Phys. Lett.} B {\bf 805} (2020) 135468.
\bibitem{rotating} R. Kumar and S. G. Ghosh, {\it JCAP} {\bf 2020} (2020) 053.


\bibitem{love1} R. A. Konoplya and A. Zhidenko, {\it Phys. Rev.} D {\bf 101} (2020) 084038.
\bibitem{love2} A. Casalino, A. Coll\'{e}aux, M. Rinaldi and S. Vicentini, {\it Phys. Dark. Univ} {\bf 31} (2021) 100770.


\bibitem{string} D. V. Singh, S. G. Ghosh and S. D. Maharaj, {\it Phys. Dark. Univ} {\bf 30} (2020) 100660.


\bibitem{rad1} S. G. Ghosh and S. D. Maharaj, {\it Phys. Dark. Univ} {\bf 30} (2020) 100687.
\bibitem{rad2} S. G. Ghosh and R. Kumar, {\it Class. Quant. Grav} {\bf 37} (2020) 245008.


\bibitem{hayward} A. Kumar and S. G. Ghosh, arXiv:2004.01131 [gr-qc].
\bibitem{bardeen} A. Kumar and R. Kumar, arXiv:2003.13104 [gr-qc].


\bibitem{ba1} S. S. Zhao and Y. Xie, {\it Eur. Phys. J.} C {\bf 77} (2017) 272.
\bibitem{ba2} X. M. Deng, {\it Eur. Phys. J.} C {\bf 80} (2020) 489.
\bibitem{ba3} B. Gao and X. M. Deng, {\it Ann. Phys} {\bf 418} (2020) 168194.
\bibitem{ba4} X. M. Deng, {\it Phys. Dark. Univ} {\bf 30} (2020) 100629.


\bibitem{w1} K. Jusufi, A. Banerjee and S. G. Ghosh, {\it Eur. Phys. J.} C {\bf 80} (2020) 698.
\bibitem{w2} P. Liu, C. Niu, X. Wang and C. Y. Zhang, arXiv:2004.14267 [gr-qc].


\bibitem{star} D. D. Doneva and S. S. Yazadjiev, arXiv:2003.10284 [gr-qc].


\bibitem{isco} M. Guo and P. C. Li, {\it Eur. Phys. J.} C {\bf 80} (2020) 588.
\bibitem{s} S. W. Wei and Y. X. Liu, arXiv:2003.07769 [gr-qc]; R. Roy and S. Chakrabarti, {\it Phys. Rev.} D {\bf 102} (2020) 024059.


\bibitem{QNM} R. A. Konoplya and A. F. Zinhailo, {\it Eur. Phys. J.} C {\bf 80} (2020) 1049; A. Arag\'{o}n, R. B\'{e}car, P. A. Gonz\'{a}lez and Y. V\'{a}squez, ibid {\bf 80} (2020) 773; S. Devi, R. Roy and S. Chakrabarti, ibid {\bf 80} (2020) 760; M. S. Churilova, {\it Phys. Dark. Univ} {\bf 31} (2021) 100748.


\bibitem{lensing} S. U. Islam, R. Kumar and S. G. Ghosh, {\it JCAP} {\bf 09} (2020) 030; R. Kumar, S. U. Islam and S. G. Ghosh, {\it Eur. Phys. J.} C {\bf 80} (2020) 1128; X. H. Jin, Y. X. Gao and D. J. Liu, {\it Int. J. Mod. Phys} D {\bf 29} (2020) 2050065.


\bibitem{stability} R. A. Konoplya and A. Zhidenko, {\it Phys. Dark. Univ} {\bf 30} (2020) 100697.


\bibitem{th1} S. A. Hosseini Mansoori, {\it Phys. Dark. Univ} {\bf 31} (2021) 100776.
\bibitem{th2}  K. Hegde, A. N. Kumara, C. L. A. Rizwan, K. M. Ajith and M. S. Ali, arXiv:2003.08778 [gr-qc].
\bibitem{th3} D. V. Singh and S. Siwach, {\it Phys. Lett.} B {\bf 808} (2020) 135658.


\bibitem{a1} C. Y. Zhang, P. C. Li and M. Guo, {\it Eur. Phys. J.} C {\bf 80} (2020) 874.
\bibitem{a2} R. A. Konoplya and A. Zhidenko, {\it Phys. Rev.} D {\bf 102} (2020) 064004.
\bibitem{a3} Y. P. Zhang, S. W. Wei and Y. X. Liu, {\it Universe} {\bf 6} (2020) 103.
\bibitem{a5} C. Liu, T. Zhu and Q. Wu, {\it Chin. Phys.} C {\bf 45} (2021) 015105.
\bibitem{a6} B. Eslam Panah, Kh. Jafarzade and S. H. Hendi, {\it Nucl. Phys. } B {\bf 961} (2020) 115269.
\bibitem{a7} X. H. Ge and S. J. Sin, {\it Eur. Phys. J.} C {\bf 80} (2020) 695.
\bibitem{a8} S. Ying, {\it Chin. Phys.} C {\bf 44} (2020) 125101.
\bibitem{a9} G. Narain and H. Q. Zhang, arXiv:2005.05183 [gr-qc].


\bibitem{Eddington} F. W. Dyson, A. S. Eddington and C. Davidson, {\it philosophical transactions of the royal society of London. series A,
containing papers of a mathematical or physical character}, {\bf 332} (1920) 291.


\bibitem{winberg} S. Weinberg, \textit{Gravitation and Cosmology}, (John Wiley and Sons, 1972)


\bibitem{Sereno} M. Sereno, {\it Phys. Rev.} D {\bf 77} (2008) 043004.


\bibitem{Will} J. Bodenner and C. M. Will, {\it Am. J. Phys} {\bf 71} (2003) 770.


\bibitem{sun} D. S. Robertson, W. E. Carter, W. H. Dillinger, {\it Nature} {\bf 349} (1991) 768; D. E. Lebach, B. E. Corey, I. I. Shapiro, M. I. Ratner, J. C. Webber,
A. E. E. Rogers, J. L. Davis, T. A. Herring, {\it Phys. Rev. Lett} {\bf 75} (1995) 1439.


\bibitem{solar} L. Amendola, C. Charmousis and S. Davis, {\it JCAP} {\bf 0710} (2007) 004.


\bibitem{PN1} C. M. Will, \textit{Theory and Experiment in Gravitational Physics}, (Cambridge University Press, Cambridge, United Kingdom, 1993).
\bibitem{PN2} C. M. Will, {\it Living Rev. Relativity} {\bf 17} (2014) 4.


\bibitem{landa} A. W. Kerr, J. C. Hauck and B. Mashhoon, {\it Class. Quant. Grav} {\bf 20} (2003) 2727.
\bibitem{data} E. Pitjeva and N. Pitjev, {\it Mon. Not. R. Astron. Soc} {\bf 432} (2013) 3431.


\end{thebibliography}
\end{document}